\documentclass[referee]{raa}            

\usepackage{graphicx}
\usepackage{natbib}
\usepackage{float}
\usepackage{subfig}
\usepackage{lineno}
\usepackage{cite}
\usepackage{underscore}

\begin{document}

   \title{A Study on Monte Carlo simulation of the radiation environment above GeV at the DAMPE orbit
}

   \volnopage{Vol.0 (200x) No.0, 000--000}      
   \setcounter{page}{1}          

 \author{S. X. Wang\inst{1,2}
    \and J. J. Zang\inst{3,1}\footnote{Corresponding author:zangjingjing@lyu.edu.cn}
    \and W. Jiang\inst{1}
    \and S. J. Lei\inst{1,2}
    \and C. N. Luo\inst{1,2}
    \and Z. L. Xu\inst{1,2}\footnote{Corresponding author:xuzl@pmo.ac.cn}
    \and J. Chang\inst{1,2}
   }

   \institute{ Key Laboratory of Dark Matter and Space Astronomy, Purple Mountain Observatory, Chinese Academy of Sciences, Nanjing 210023, China\\
\and
School of Astronomy and Space Science, University of Science and Technology of China, Hefei 230026, China\\
\and
School of Physics and Electronic Engineering, Linyi University, Linyi 276000, China\\
   }
   \date{Received~~2009 month day; accepted~~2009~~month day}

\abstract{The Dark Matter Particle Explorer (DAMPE) has been undergoing a stable on-orbit operation for more than 6 years and acquired observation of over 11 billion events. And a better understanding of the overall radiation environment on the DAMPE orbit is crucial for both simulation data production and flight data analysis. In this work, we study the radiation environment at the low Earth orbit and develop a simulation software package using the framework of ATMNC3, in which state-of-the-art full 3D models of the Earth's atmospheric and magnetic-field configurations is integrated. We consider in our Monte Carlo procedure event-by-event propagation of the cosmic rays in the geomagnetic field and their interaction with the Earth's atmosphere, focusing on the particles above GeV that are able to trigger the DAMPE data acquisition system. We compare the simulation results with the cosmic-ray electrons and positrons (CREs) flux measurements made by DAMPE. The overall agreement on both the spectral and angular distribution of the CREs flux demonstrates that our simulation is well established. Our software package could be of more general usage for the simulation of the radiation environment at the low Earth orbit of various altitudes.
}
   \authorrunning{S.-X. Wang et al. }            
   \titlerunning{Study on Monte Carlo simulation of radiation environment}  

   \maketitle

%
%
\section{Introduction}           
\label{sect:intro}
The Dark Matter Particle Explorer (DAMPE), launched on Dec. 17, 2015, is a space-borne high-energy particle detector covering an energy range from GeV to TeV (\citealt{CHANG20176,Xu2018,HuangYY2020}). The observation results of DAMPE can be used for scientific research on the origin of cosmic rays (\citealt{proton2019,Yuan2020,Yuan2018}), indirect detection of dark matter (\citealt{Nature2017,Yuan2017,Xu2021,DAMPE2021hsz}) and gamma-ray astronomy (\citealt{Duan2019}). DAMPE has been operating on a solar synchronized orbit of 500~km for more than 6 years and has acquired observational data of over 11 billion cosmic-ray events. At this low Earth orbit (LEO\footnote{https://www.sciencedirect.com/topics/earth-and-planetary-sciences/low-earth-orbit}), massive data are collected in a complex cosmic-ray radiation environment mainly formed by the interaction of the galactic cosmic rays (GCRs) with the geomagnetic field and the Earth's atmosphere. The charged GCRs propagating in interplanetary space are modulated by solar activities, showing a flux fluctuation with a period of about 11 years. The transportation of charged GCRs in the Earth magnetosphere region is mainly affected by the geomagnetic field, forming several phenomena well known, such as the radiation belt, the South Atlantic Anomaly area, and the East-West effect. At even closer distance to the Earth, GCRs may interact with atoms in the atmosphere, generating a large amount of secondary particles. As DAMPE keeps operating in such a complex radiation environment, accurate on-orbit simulation of the radiation environment is the basis for the fine calibrations of the detector that is crucial to understand the flight data, such as the minimum ionized particles (MIPs) calibration, the geomagnetic rigidity cutoff measurement, the event rate and data volume estimation of the data acquisition (DAQ) system, and so on.

Given the critical role played by a correct understanding of the radiation environment at the DAMPE orbit in both the simulation data production and flight data analysis, we study in this work the Monte Carlo (MC) simulation of the radiation environment at the LEO, and the DAMPE orbit in particular. As the inputs of our simulation, we make use of the fluxes of primary cosmic-ray protons, helium nuclei, electrons and positrons (CREs) above GeV measured by the AMS02 (\citealt{PhysRevLett.102, PhysRevLett.101, PhysRevLett.103}). We then back trace the propagation of these charged particles in the geomagnetic field, obtaining the distribution of primary cosmic-ray particles on the satellite orbital plane. The interaction between these primary cosmic-ray particles and the Earth's upper atmosphere is simulated to generate secondary particles. All these primary and secondary particles are then connected to the DAMPE official Geant4 package to simulate their interaction with the detectors, and finally produce reconstructed simulation data in the same format as the flight data. Our simulation of the radiation environment focus on the high-energy GCRs above GeV, as the solar wind particles below hundreds of MeV can not trigger the DAQ system (\citealt{Zhang2019}). To investigate the accuracy of our on-orbit simulation, several validations on the spectral and spatial distribution are performed in several different radiation regions. Our results show good agreement between flight data and simulation.

\section{Simulation methods}
\begin{figure*}[!ht]
  \centering
  \includegraphics[scale=0.21]{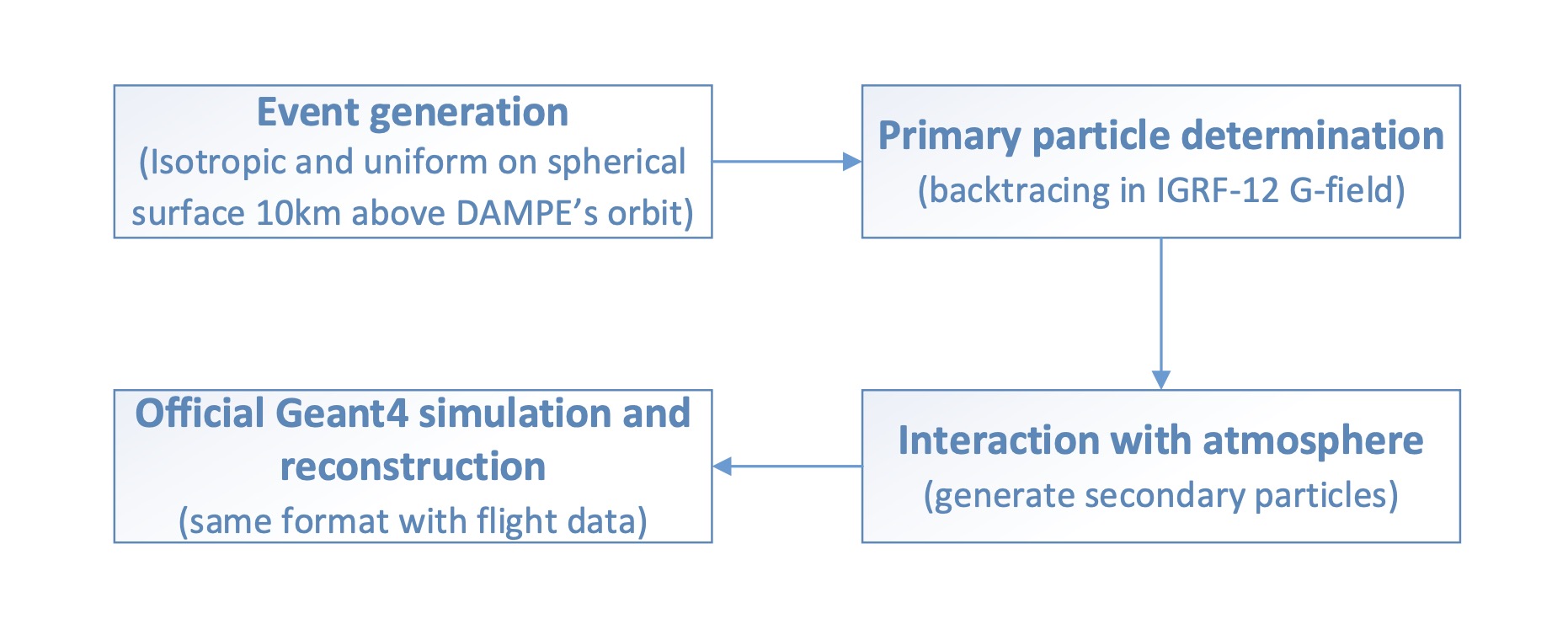}
  \caption{The procedure of on-orbit simulation. First, the GCR events are generated isotropically and uniformly on an Earth-centered spherical surface at an altitude of 510~km. Then a time-reversed trajectory backtracing of the particle in a magnetic field described by the IGRF-12 model determines whether it is a primary or unreasonable one. In a third step, the primary particles interact with the atmosphere to generate secondary particles. And at last, the simulated primary and secondary particles are imported into the DAMPE official Geant4 simulation and reconstruction package to produce simulation data in the same format as flight data.}
  \label{Fig:Simu-Flow}
\end{figure*}

\begin{figure*}[!ht]
  \centering
  \includegraphics[scale=0.38]{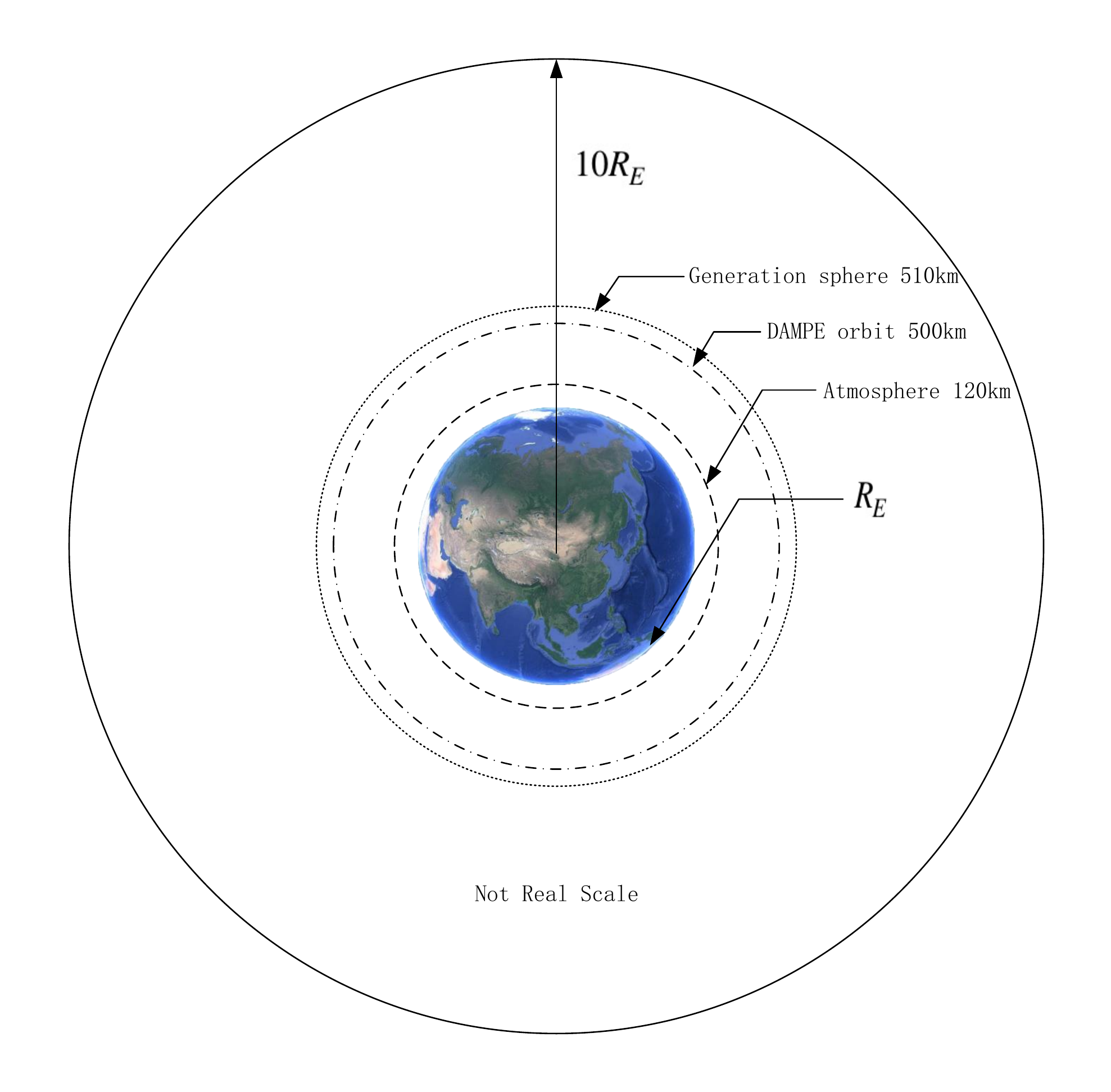}
  \caption{A schematic of the setup of our simulation of the radiation environment at the DAMPE orbit. Particles are uniformly generated on the Generation sphere 10~km above the DAMPE orbit at 500~km. The back-tracing method enables the selection of primary particles coming from beyond 10 Earth-radii, and whose interaction with the Atmosphere is further simulated to produce the secondaries.}
  \label{Fig:Sketch}
\end{figure*}

The ultimate goal of the software is to generate simulation data in the same format as the flight data by simulating the cosmic-ray environment in the DAMPE orbit and the process of flight data acquisition. For this purpose, the on-orbit simulation is separated into four steps as illustrated in Figure \ref{Fig:Simu-Flow}: the event generation, the primary particle determination, the secondary particle generation, and the official Geant4 simulation. The first three steps are developed based on the ATMNC3 (\citealt{PhysRevD.70.043008}) software framework that is originally developed to calculate atmospheric neutrinos flux, which but can also be used to simulate the cosmic-ray flux near the Earth with proper modification. In our simulation, the GCRs events are uniformly generated on an Earth-centered spherical surface (10~km above the DAMPE orbit, as shown in Figure \ref{Fig:Sketch}) with an isotropic flux, and then only inward going particles are considered to ensure that they can reach the DAMPE orbit. A so called back tracing process is then employed to determine if the particle is reasonable. For each event, the charge sign and the momentum direction are reversed, and the particle will propagate along its time-reversed trajectory in the geomagnetic field. If the particle reaches a distance of 10 Earth-radii, it is considered a reasonable one, otherwise a forbidden trajectory or a primary particle has been counted.
During the backtracing, the tracing step length is optimized to balance CPU time consumption and trajectory accuracy. As the inputs of our simulation, we use the measurements of the AMS02 for the incoming fluxes of various GCR components, and only three GCR components of the most interest are considered, namely, protons (\citealt{PhysRevLett.103}), helium nuclei (\citealt{PhysRevLett.101}), and CREs (\citealt{PhysRevLett.102}). We consider in our simulation a full 3D description of the Earth's atmospheric (NRLMSISE00 atmospheric model, \citealt{https://doi.org/10.1029/2002JA009430}) and magnetic-field (IGRF-12 geomagnetic field model, \citealt{nerc511731}) configurations using the latest available models. The NRLMSISE00 model describes the atmospheric density, composition and changes over time from sea level to 1000~km altitude. The IGRF12 is an international geomagnetic reference field model provided by the V-MOD International Cooperative Group of the International Association for Geomagnetism and High Altitude Atmospheres (IAGA), covering the period of 2015 - 2020. The relative error of IGRF12 is about 1~nT while the typical field strength is about 40000~nT. Event-by-event simulation of the GCRs propagation in the geomagnetic field and their interaction with the Earth's atmosphere is implemented in the ATMNC3 framework also with the help of the DMPJET3 (\citealt{RANFT2003392}) and PHITS (\citealt{NIITA20061080}) particle interaction models. The DMPJET3 is a new version of the Monte Carlo (MC) event generator DPMJET for air shower simulation, and the PHITS is a MC particle propagation model that applies to particles in the energy range from 1 GeV to 200 GeV.

After the ATMNC3 simulation, for the validation of our simulation results by comparing to flight data, all the primary and secondary particles generated are imported into the DAMPE official Geant4 simulation and reconstruction package (Jiang et al. 2020), where the relativistic inelastic and elastic interaction between a particle and the DAMPE detector is simulated. In the end, the MC data are reconstructed using the same algorithm for the flight data reconstruction, and a simulation data sample in the same format as flight data are generated.

\section{Validation of on-orbit simulation}
In this section, we validate our on-orbit simulation by comparing the MC results with the CREs flux measurement of DAMPE. The primary CREs nearly isotropically propagating in interplanetary space, however, are deflected by geomagnetic force when entering the Earth's magnetic field. As a consequence, the energy spectrum of primary CREs near the Earth is no longer a single power-law but a broken power-law with a magnetic latitude dependent cutoff, the so called geomagnetic rigidity cutoff. And the isotropic angular distribution is also broken by the rigidity cutoff and the so the called East-West asymmetry in the angular distribution is formed. Meanwhile, some secondary CREs generated in top-layer atmosphere can also trigger the DAQ. So the CREs observed by the DAMPE have two origins, one population is from the primary GCRs and the other comes from the interaction between GCRs and atmosphere. By comparing the spectrum of primary and secondary combined CREs, we can validate both the determination method of the primary particles and their interaction with the Earth's atmosphere.

\subsection{Primary and secondary CREs flux with geomagnetic rigidity cutoff}
In the analysis, we select primary and secondary CREs flux calculated in the magnetic equator region where the geomagnetic field lines are almost parallel to the surface of the Earth, the charged particles are deflected by the Lorentz force, while in the polar region, the magnetic field lines are open and charged cosmic-ray can reach the Earth’s surface along field lines. As a consequence, the geomagnetic rigidity cutoff is much higher in the equatorial region, where the typical value is about a few tens of GeV, well in the sensitive energy range of the DAMPE. For comparison, we use the CREs fluxes near the geomagnetic equator in DAMPE orbit (\citealt{inproceedings}), and the DAMPE measurement of the CREs fluxes above 1.2 times the rigidity cutoff is normalized to the fluxes published by AMS02(\citealt{PhysRevLett.102}).

\begin{figure*}[!ht]
  \centering
  \includegraphics[scale=0.7]{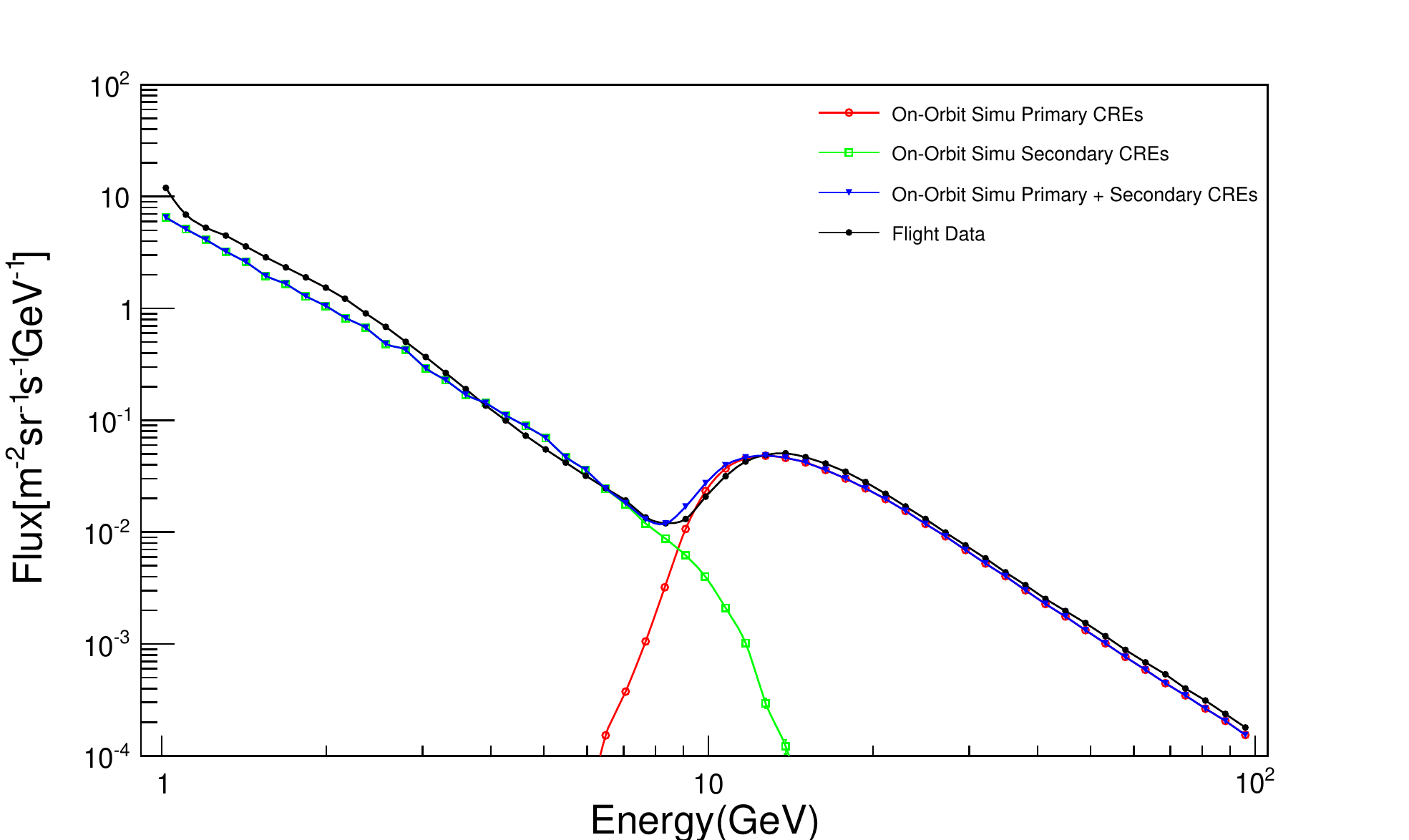}
  \caption{The CREs fluxes of the MC and flight data (reweighted to AMS02) (\citealt{PhysRevLett.102}) are compared in McIlwain L(\citealt{Hilton1971}) range of 1-1.14. The flight data CREs flux is normalized to the simulated flux above 20 GeV, in that energy range the primary CREs flux is almost independent to the geomagnetic field.}
  \label{Fig:CREs-Flux}
\end{figure*}

As show in Figure \ref{Fig:CREs-Flux}, the CREs flux is broken into two parts due to the rigidity cutoff. Above the cutoff, primary CREs dominate the population, while below the cutoff they are almost all secondary ones since few primary CREs with that energy can reach the equator region due to the existence of the Earth and the geomagnetic field. By comparing the spectra of the flight data and MC data, the spectra validation has been performed in all region of DAMPE orbit. The primary and secondary flux of simulated CREs agrees well with flight data especially near the rigidity cutoff region where the contributions from neither primary nor secondary are ignorable. The overall agreement demonstrates that the spectra of both primary and secondary CREs have been well described.

\subsection{Angular distribution of primary and secondary CREs}
\begin{figure*}[!ht]
  \centering
  \subfloat[]{
    \includegraphics[width=.5\textwidth,height=35mm]{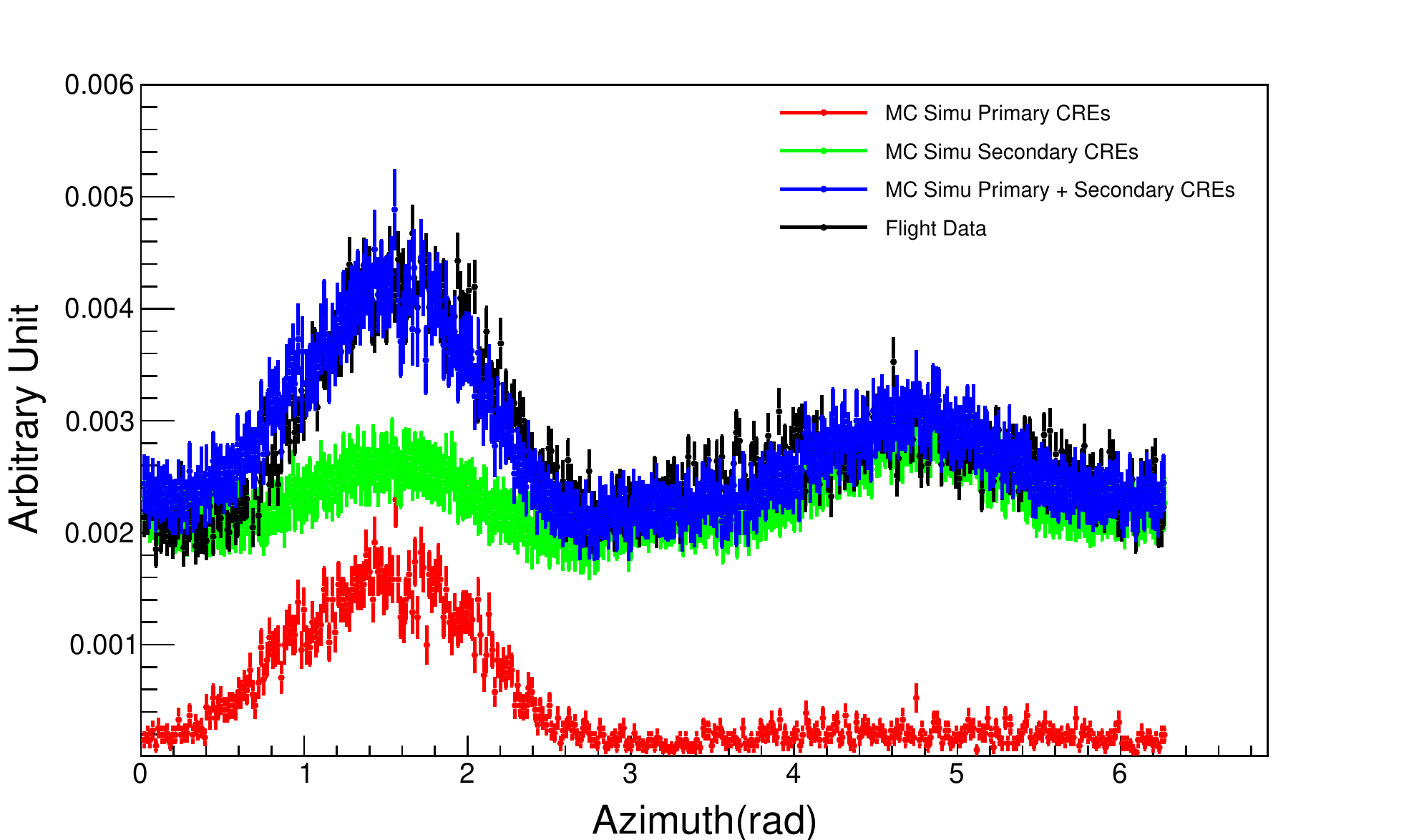}
  }
  \subfloat[]{
    \includegraphics[width=.5\textwidth,height=35mm]{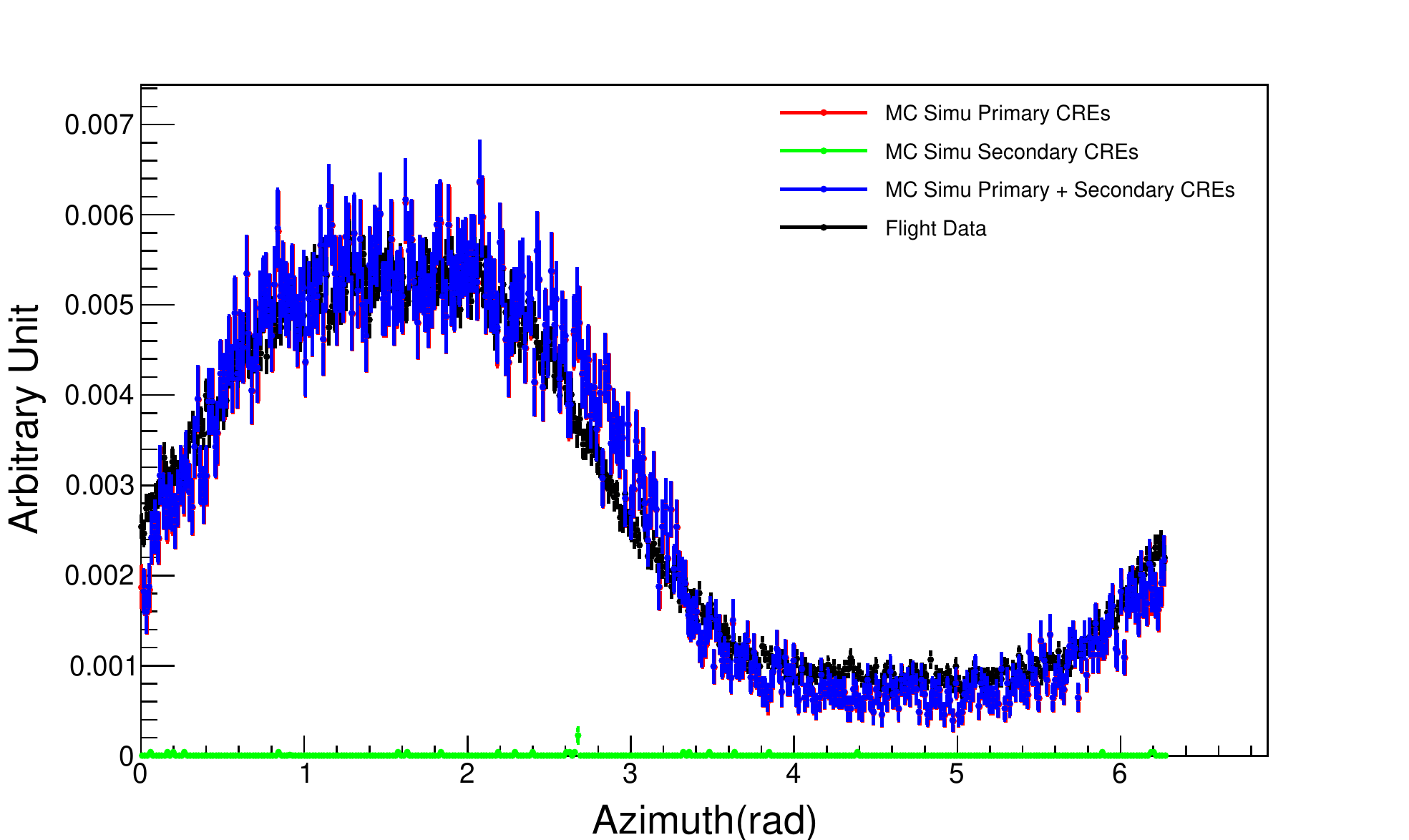}
  }
  \\
  \subfloat[]{
    \includegraphics[width=.5\textwidth,height=35mm]{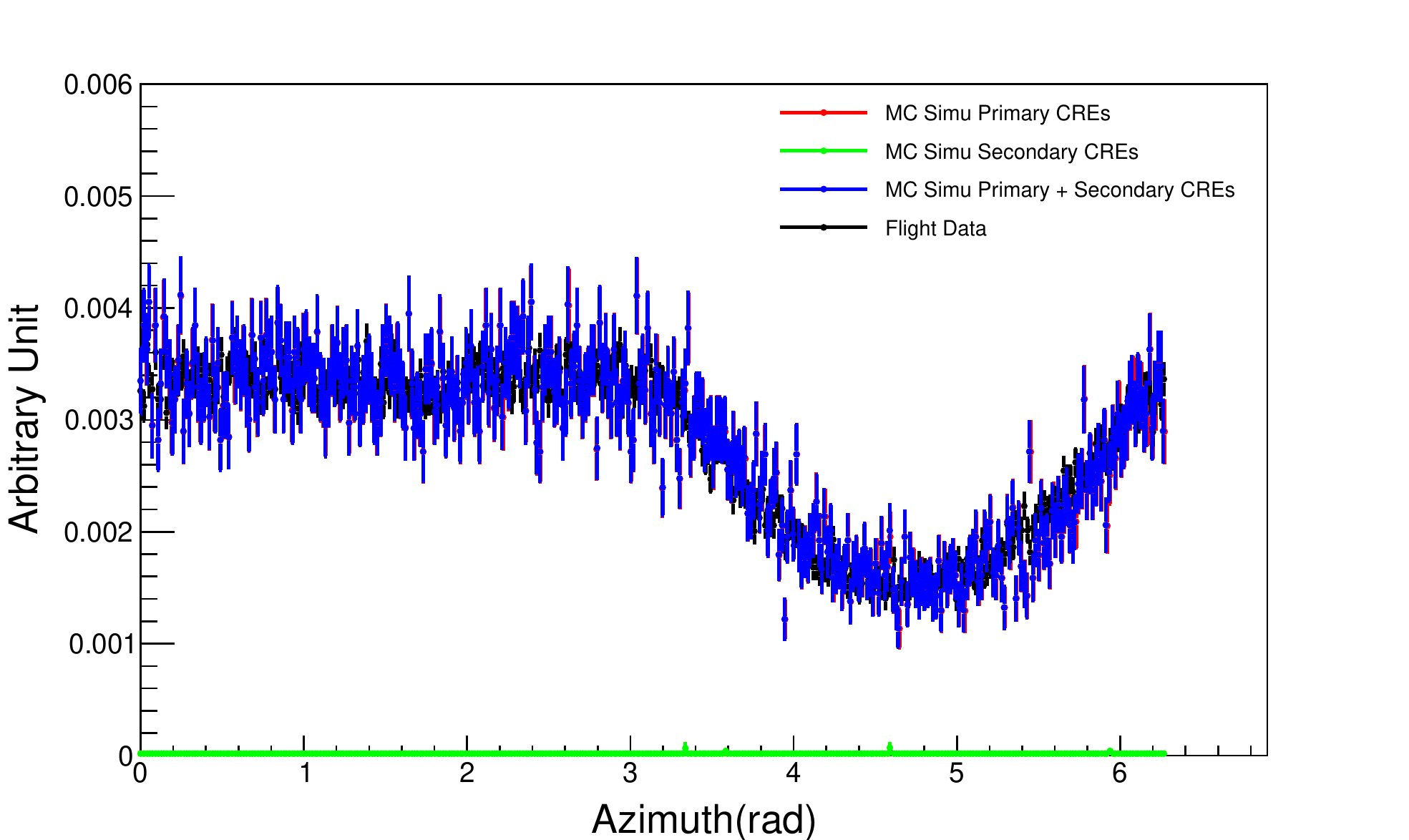}
  }
  \subfloat[]{
    \includegraphics[width=.5\textwidth,height=35mm]{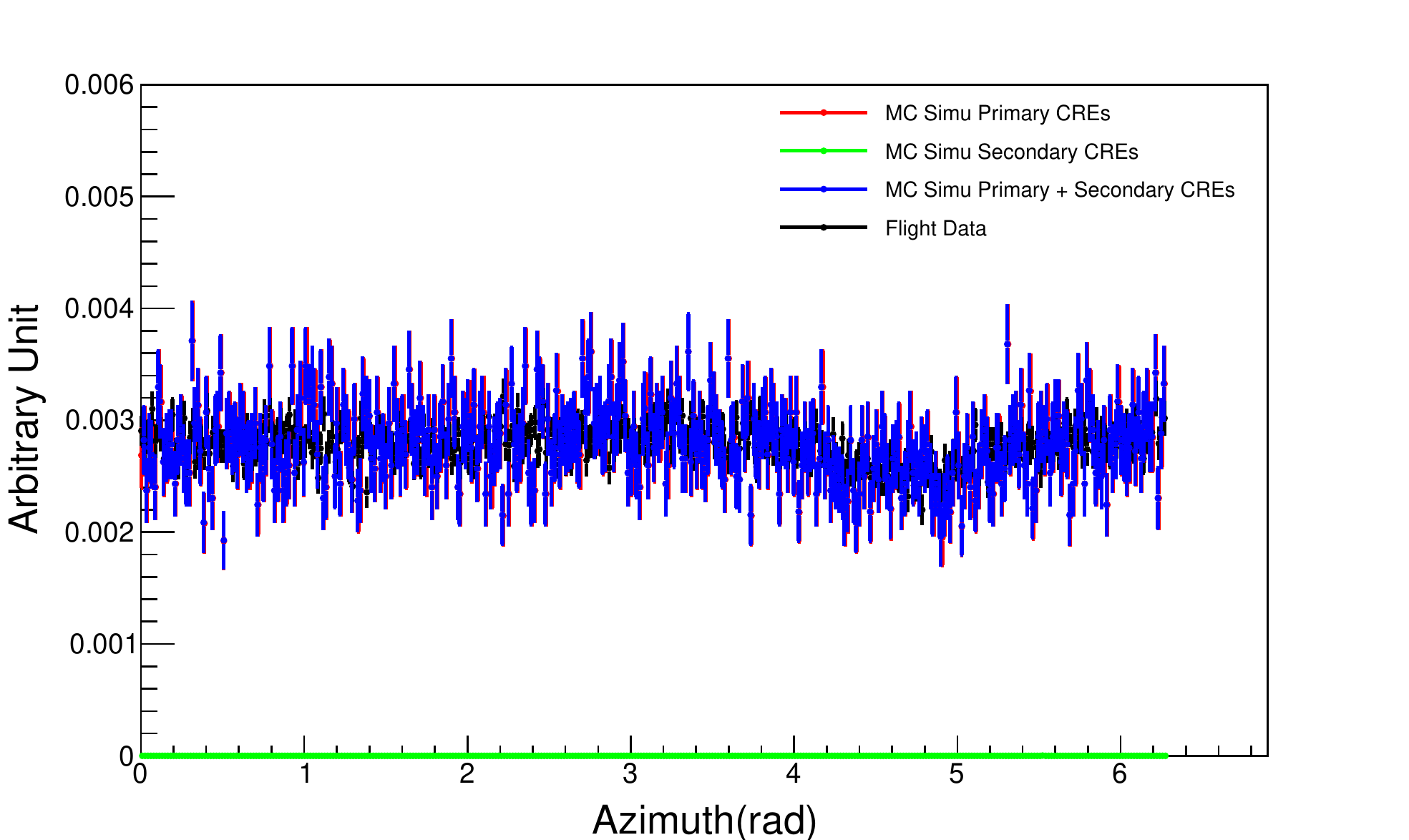}
  }

  \caption{The CREs azimuthal distribution of MC and flight data are compared for different energy ranges of (a) [8.0, 10.3] GeV , (b) [12.1, 13.2] GeV , (c) [14.4, 15.7] GeV, and (d) [17.1, 18.6] GeV in the McIlwain L value range of [1, 1.14]. The CREs azimuthal distribution is averaged over all zenith angle. The flight data CREs azimuthal distribution is from Ref (\citealt{inproceedings}). Quite good agreements between flight data and our simulation could be seen in the figures for the CREs azimuthal distribution in different energy ranges.}
  \label{Fig:Azimuth}
\end{figure*}

\begin{figure*}[!ht]
  \centering
  \includegraphics[scale=0.7]{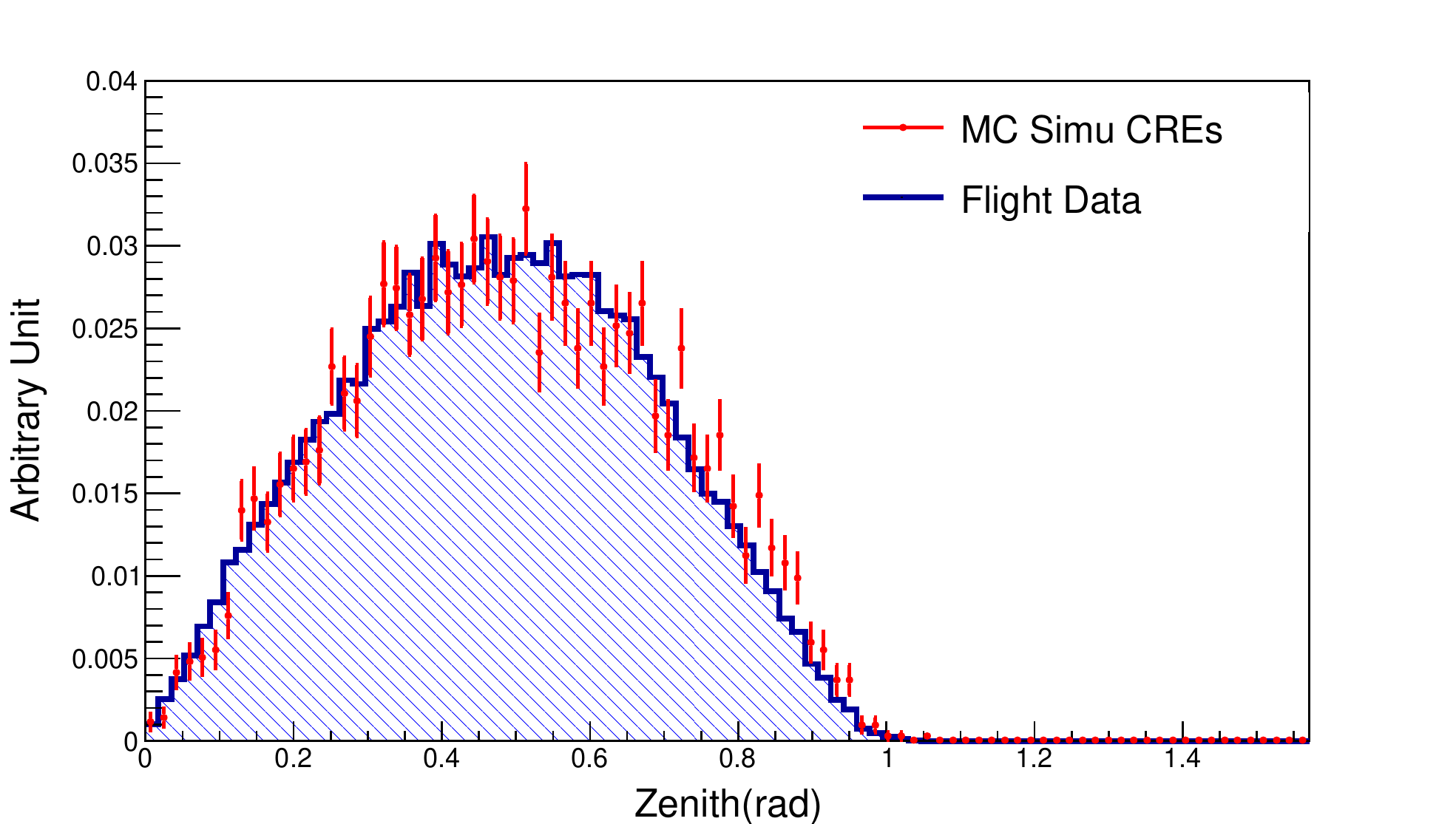}
 \caption{The CREs Zenithal distributions of MC and flight data and on-orbit are compared. The flight data result here takes account of all the CREs observed by DAMPE, and the MC result is convolved with the DAMPE acceptance. An overall good agreement between flight data and simulation is also seen for the CREs zenithal simulation.}
  \label{Fig:Zenith}
\end{figure*}

As mentioned above, the charged GCRs propagating in the interplanetary space show highly isotropic in their direction of arrival. Deflected by the Lorentz force while propagating in the geomagnetic field, they are no longer isotropic on LEO, but show asymmetric angular distribution, i.e., the so-called East-West effect. That is, more positively charged particles are seen from west moving eastward, and vice versa for negatively charged particles. To describe the asymmetry angular distribution, the azimuth and zenith angle is defined in the Earth-centered coordinate system. In the system, the coordinate origin is in the center of the earth, the azimuth angle is defined as geographical azimuth, particles from North have an azimuth of $0^{\circ}$, and East is at $90^{\circ}$, South is at $180^{\circ}$, West is at $270^{\circ}$. Figure \ref{Fig:Azimuth} shows the azimuthal distributions of CREs extracted from flight data of DAMPE and on-orbit simulation in four typical energy bins in an L interval of 1-1.14. Since DAMPE has no ability to distinguish the origins of CREs, flight data only gives the entire azimuth distribution including both primary and secondary. Our on-orbit simulation can trace and distinguish primary and secondary CREs. The overall good agreement of azimuth between flight data and primary+secondary demonstrates that the angular distribution has been described well. Benefit from this structured azimuth distribution, the method of estimating the fraction of primary part is developed, and that is the main idea of simultaneous measurement flux of primary and secondary CREs.
In the Earth-centred coordinate system, zenith is defined as the angle between the particle motion direction and the line pointing from the satellite to the Earth center. Figure \ref{Fig:Zenith} shows the zenithal direction for CREs collected on DAMPE orbit with an L range of $1-1.14$. The distribution is convolved with the DAMPE acceptance. There is no significant discrepancy between real data and simulation demonstrating zenith direction has been well described too.

\section{Conclusion}
A software package for the DAMPE on-orbit radiation environment simulation is developed using the ATMNC3 framework, within which state-of-the-art models of the geomagnetic field and the atmosphere are also integrated. By taking delicate consideration of the GCRs propagation in the geomagnetic field and their interaction with the atmosphere, the software package is able to simulate the cosmic-ray radiation environment on the DAMPE orbit with sufficient accuracy and reasonable computational consumption. The simulation results are validated by comparing to the real observations for the CREs flux in several energies and geomagnetic latitude ranges. The overall agreements on the spectral and angular distributions demonstrate that the CREs radiation environment is well simulated. Specifically, the structured azimuthal distribution can severe as a tool to estimate the fraction of primary or secondary CREs. Our simulation tool thus could be helpful to the scientists and engineers working on the data analysis of space detectors after launch as well as in their designation of science satellite before launch. The software package can also be used to simulate the radiation environment in other LEOs by simply changing the orbital altitude, so it can also be used in possible space science projects in the future, such as HERD(\citealt{HERD2014}) and VLAST\footnote{https://m.gmw.cn/baijia/2020-10/15/1301673634.html}.

\begin{acknowledgements}
This work is supported in part by the National Key R\&D Program of China (2021YFA0718404), the National Natural Science Foundation of China(Nos. 11773085, U1738207, 12173098), the Youth Innovation Promotion Association CAS, the Scientific Instrument Developing Project of the Chinese Academy of Sciences, Grant No. GJJSTD20210009.

\end{acknowledgements}

\label{lastpage}

\bibliographystyle{raa}

\end{document}